\documentclass[pra,aps,showpacs,twocolumn]{revtex4}

\usepackage{epsf}
\usepackage{amsmath}
\usepackage[dvips]{color}

\newcommand {\be}{\begin{equation}}
\newcommand {\ee}{\end{equation}}
\newcommand {\bea}{\begin{eqnarray}}
\newcommand {\eea}{\end{eqnarray}}

\newcommand {\mrm}{\mathrm}
\newcommand {\dd} {\mathrm{d}}
\newcommand {\sn} {\mathrm{sn}}
\newcommand {\ns} {\mathrm{ns}}
\newcommand {\dn} {\mathrm{dn}}
\newcommand {\cs} {\mathrm{cs}}
\newcommand {\cn} {\mathrm{cn}}
\newcommand {\ds} {\mathrm{ds}}
\newcommand {\sech} {\mathrm{sech}}
\newcommand {\csch} {\mathrm{csch}}

\begin{document}


\title{Effect of a potential step or impurity on the Bose-Einstein condensate mean field}
\author{B. T.  Seaman, L. D. Carr, and M. J. Holland}
\affiliation{JILA, National Institute of Standards and Technology
and Department of Physics,
\\University of Colorado, Boulder CO  80309-0440, USA}
\date{\today}


\begin{abstract}
The full set of stationary states of the mean field of a Bose-Einstein condensate in the presence of a potential step or point-like impurity are presented in closed analytic form.  The nonlinear Schr\"odinger equation in one dimension is taken as a model.  The nonlinear analogs of the continuum of stationary scattering states, as well as evanescent waves, are discussed. The solutions include asymmetric soliton trains and other wavefunctions of novel form, such as a pair of dark solitons bound by an impurity.

\end{abstract}

\pacs{03.75.Hh, 05.30.Jp, 05.45.Yv}

\maketitle


\section{Introduction}
\label{sec:intro}

The nonlinear Schr\"odinger equation (NLS) models many kinds of wave phenomena.  The NLS appears in diverse fields such as nonlinear optics~\cite{Talanov1965}, gravity waves on deep water~\cite{Hasimoto1972}, magneto-static spin waves \cite{Scott2000}, solitons in liquid crystals~\cite{Conti2003}, and magneto-sonic solitons in the atmospheric magneto-pause boundary layer~\cite{Stasiewicz2003}.  It also describes the dynamics of the mean field of a weakly interacting atomic or molecular Bose-Einstein condensate (BEC)~\cite{Anderson1995,Davis1995,Bradley1995,Bradley1997}, where it is known as the Gross-Pitaevskii equation~\cite{Gross1961,Pitaevskii1961}.

Here, we consider the steady state response of the mean field of a
BEC to a potential step or a delta function potential, as modeled
by the NLS.  These potentials may be easily produced in present
experiments on the BEC. The former may be realized by a detuned
laser beam shined over a razor edge to make a sharp barrier, where
the diffraction-limited fall-off of the laser intensity is smaller
than the healing length of the condensate, so that the potential
is effectively a step function.  The latter models the response of
the condensate to an impurity of a length scale smaller than the
healing length, which could be realized by a tightly focused laser
beam, by another spin state of the same atom, or by any other
object, as for instance another alkali atom, confined in an
optical trap.  Moreover, the solution methods developed in this
article may be generalized to arbitrarily complicated piecewise
constant potentials.

The complete set of stationary solutions to the NLS with a constant potential
on the infinite line were discovered by Zakharov and
Shabat~\cite{Zakharov1971,Zakharov1973}.  The stationary solutions of the NLS under periodic
and box boundary conditions has also been solved
analytically~\cite{Carr2000,Carr2000_2}, as well as the finite
well~\cite{Carr2001_2}.  The parabolic potential has been solved
numerically~\cite{Kivshar2001}.  The potential step has been
examined theoretically and experimentally for the linear
Schr\"odinger equation with a
constant~\cite{Villavicencio2002,Muga2000} and
oscillating~\cite{Felber1996} step.  Klein examined similar
problems experimentally by deflecting neutron beams with a
vibrating crystal~\cite{Klein1967}.  In addition, symmetric
steady-state solutions with a point-like impurity potential have
been studied by Hakim~\cite{Hakim1997} and
Taras-Semchuk~\cite{Taras1999} in certain limiting cases.
Point-like impurity potentials have been studied extensively, such
as with helium impurities in a BEC~\cite{Capuzzi2000,Harms1998},
BEC formation initiated by point like
impurities~\cite{Cavalcanti2002} and impurity scattering in a BEC
of sodium~\cite{Chikkatur2000}.  The superfluid transmission of matter waves across various potentials has been studied~\cite{Leboeuf2003,Pavloff2002}.  In addition, bound solutions to
the NLS are the one-dimensional analog to the pinned vortex
solutions which occur when a discontinuity is present in a two
dimensional system, such as in a two dimensional high-$T_c$
superconducting system.

In order to obtain the full set of stationary states in closed
analytic form, we assume the BEC to be in the
quasi-one-dimensional regime.  When the transverse dimensions of
the BEC are on the order of its healing length, and its
longitudinal dimension is much longer than its transverse ones,
the 1D limit of the 3D NLS is appropriate to describe the
system~\cite{Carr2000_3}.  The 1D NLS, with an external potential,
$V(x)$, may be written \be
i\partial_{t}\Psi=-\frac{1}{2}\partial_{xx}\Psi+g
|\Psi|^{2}\Psi+V(x)\Psi\, , \label{eqn:NLSE} \ee where a harmonic
oscillator confinement in the transverse directions with frequency
$\omega$ has been assumed~\cite{Olshanii1998} for atoms of mass M,
the length has been rescaled according to units of the oscillator
length, $l_{ho}=(\hbar/M\omega)^{1/2}$, and energy rescaled
according to units of the oscillator energy, $\hbar \omega$.  The
renormalized 1D coupling, $g \equiv 2 a_s$, where $a_s$ is the
$s$-wave scattering length, characterizes the short-ranged
pairwise interactions between atoms.  The wave function or order
parameter $\Psi(x,t)$ has the physical meaning of
$\Psi(x,t)=\sqrt{\rho(x,t)}\exp[i\phi(x,t)]$, where $\rho(x,t)$ is
the longitudinal line density and the longitudinal superfluid
velocity is given by $v(x,t)=\partial \phi(x,t)/\partial x$.  Both
attractive and repulsive atomic interactions, {\it i.e.}, $g>0$
and $g<0$, shall be considered.

In the case where the harmonic oscillator length approaches the
$s$-wave scattering length, $l_{ho}\simeq a_s$, the 1D NLS no
longer models the system and a one-dimensional field theory with
the appropriate effective coupling constant must be considered
instead~\cite{Olshanii1998}. Since $a_s$ is on the order of
angstroms for typical BEC's, this regime is not relevant to the
present study.  Thus, it should be noted that this study does not
examine the Tonks-Girardeau regime~\cite{Girardeau1960}, where
quantum fluctuations become important and the Gross-Pitaevskii
equation no longer models the system.

With the  experimental demonstration of Feshbach resonances in
BEC's of dilute atomic gases~\cite{Roberts1998,Inouye1998}, it is
possible to alter the $s$-wave scattering length and, hence, the
nonlinearity of the NLS.  Near a Feshbach resonance, the
scattering length becomes a function of a uniform background
magnetic field.  By altering the magnetic field, the effects of
the nonlinearity can be experimentally controlled.  It is
therefore important to be able to characterize the complete set of
solutions as a function of the interaction strength.

This article is organized as follows.  In Sec.~\ref{sec:steady},
the full set of stationary solutions of the NLS along the infinite
line are presented.  In Sec.~\ref{sec:Step}, the stationary
solutions influenced by a step potential are discussed.  The
effects of a point-like impurity are presented in
Sec.~\ref{sec:Delta}.  Section~\ref{sec:linear} shows the
connection between the NLS solutions and the well-known solutions
to the linear Schr\"odinger equation.   The results are discussed
in Sec.~\ref{sec:conclusion}, including a physical interpretation
of all solution variables.  In App.~\ref{app:jacobi}, the special
functions used throughout the article are briefly reviewed.
Finally, in App.~\ref{app:solution} the fact that the solutions
discussed in Sec.~\ref{sec:steady} are indeed the full set of
stationary solutions to the 1D NLS with a constant
potential~\cite{Carr2000,Carr2000_2} is demonstrated formally.


\section{Constant Potential}
\label{sec:steady}

A brief review is given of the stationary solutions of
Eq.~(\ref{eqn:NLSE}) of the form,
 \be
\Psi(x,t)=R(x)\exp[i\phi(x)]\exp[-i\mu t]\, , \ee where $\mu$ is
the eigenvalue.  There are several excellent references which consider stationary solutions in a constant potential~\cite{Infeld2000,Kivshar1998,Carr2000_3,Infeld1979,Yuen1975}.  Assuming a constant external potential $V$, the
separation of Eq.~(\ref{eqn:NLSE}) into real and imaginary parts
gives \bea R\, \partial_{xx} \phi+2 (\partial_x R)\,( \partial_x
\phi)&=&0\, , \label{eqn:fluidConservation}
\\
\frac{1}{2} \partial_{xx} R+\frac{1}{2} (\partial_x \phi)^2\, R +
g\, R^3 + V\, R&=&\mu\, R\, . \label{eqn:RxxPhi} \eea
Equation~(\ref{eqn:fluidConservation}) can be integrated to give
\be
\partial_x \phi  =  \frac{\alpha}{\rho}\, ,
\label{eqn:phix} \ee where $\alpha$ is an undetermined constant of
integration, and $\rho(x)=R(x)^2$ is the single-particle density.
Substituting Eq.~(\ref{eqn:phix}) into Eq.~(\ref{eqn:RxxPhi}) and
integrating yields \be \frac{\dd \rho}{\dd x}=2\sqrt{g\,
\rho^3-2(\mu-V)\rho^2+C\,\rho-\alpha^2}\, . \label{eqn:drho} \ee
Integrating a second time gives \be \int \frac{1}{\sqrt{g
\rho^3-2(\mu-V) \rho^2+C \rho-\alpha^2}}\, \mathrm{d}\rho=2x+x_0\,
, \label{eqn:rhoInt} \ee where $C$ and $x_0$ are undetermined
constants of integration.  It is shown in App.~\ref{app:solution}
that the only solutions to this integral equation are given by the
Jacobian elliptic functions~\cite{Bowman1961,Abramowitz1964}.  In
App.~\ref{app:jacobi} these special functions are briefly
reviewed.  The most general form of the solution is then given by
\be \rho=A\, \sn^2(b\,x+x_0,k)+B\, , \label{eqn:sn} \ee where $sn$
is one of the Jacobian elliptic functions, $A$ is a density
prefactor, $b$ is a translational scaling, $x_0$ is a
translational offest, $k$ is the elliptic parameter, and $B$ is a
vertical density offset.  The period of the density is given by
$2\, K(k)/b$, where $K(k)$ is the complete elliptic integral of
the first kind.  Since the density is positive definite, the
variables are restricted such that $B\ge 0$ and $B+A\ge 0$.  In
Sec.~\ref{sec:conclusion} the relation between these variables and
the mean number density, energy density, and momentum density are
calculated and discussed for the nonlinear wave of
Eq.~(\ref{eqn:sn}).

It should be noted that since many of the solutions are unbounded,
the norm of the wavefunction remains unconstrained.  This is in
contrast to bound or localized solutions when the normalization,
\be \int_{-\infty}^{+\infty}(\rho-\overline{\rho})\, \dd x = 1\, ,
\label{eqn:norm} \ee may be used \cite{Kivshar1998}, where
$\overline{\rho}$ is the average density.  Alternatively, it is
possible to use a nonlinear scaling of the position and density,
$x \rightarrow a x$ and $\rho \rightarrow a^{-2} \rho$, to scale
the maximum density on one side of the boundary to unity. However,
throughout this paper the wavefunction remains unnormalized.

One may then determine the variables $\mu$, $\alpha$, and $k$, in
terms of $A$, $B$, $b$, and $g$ by substituting Eq.~(\ref{eqn:sn})
into Eq.~(\ref{eqn:drho}) and equating powers of the elliptic $sn$
function to give \bea \mu&=&\frac{1}{2}(b^2+(A+3B)g)+V\, ,
\label{eqn:mu}
\\
\alpha^2&=&B(A+B)(b^2+B g)\, , \label{eqn:alpha}
\\
k^2&=&\frac{A}{b^2}g\, . \label{eqn:m} \eea This leaves the
eigenvalue, $\mu$, the constant of integration of the phase,
$\alpha$, and the elliptic parameter, $k$, determined up to $A$,
$B$, $b$, and the interaction strength, $g$.  Note that the fact
that $\alpha$ enters into the equations only as $\alpha^2$ implies
that all {\it nontrivial phase} solutions, {\it i.e.}, those for
which $\alpha \neq 0$, are doubly degenerate, as $\pm \alpha$ lead
to the same value of the eigenvalue, $\mu$, without otherwise
changing the form of the density or phase.  We shall use the term
{\it trivial phase} to refer to solutions for which the phase is
spatially constant.

In the following two sections these results are applied to
piecewise constant potentials.  In particular, the potential step
and the delta function potential are examined.


\section{Potential Step}
\label{sec:Step}

In this section the complete set of solutions to the NLS with an
external step potential of height $V_0$ beginning at $x = 0$, \be
V(x)=V_0 \, \theta(x)\, , \ee are presented analytically, given
the solution parameters on the negative $x$, or left, side of the
step.  In the following two subsections, the general solution to
the NLS with a potential step and some particular examples are
discussed.

\subsection{General Solution}
\label{subsec:StepGeneral}

Applying the two boundary conditions of continuity of the wave
function and continuity of the derivative of the wave function
gives the following five conditions, \bea \rho(0^+)&=&\rho(0^-)\,
, \label{eqn:rhoCont}
\\
\partial_x \rho(0^+)&=&\partial_x \rho(0^-)\, ,
\label{eqn:rhoxContS}
\\
\phi(0^+)&=&\phi(0^-)+2 \pi n\, , \label{eqn:phiCont}
\\
\alpha(0^+)&=&\alpha(0^-)\, , \label{eqn:alphaCont}
\\
\mu(0^+)&=&\mu(0^-)\, , \label{eqn:muCont} \eea where $n$ is an
integer.  The first two conditions represent the continuity of the
density, Eq.~(\ref{eqn:rhoCont}), and the derivative of the
density, Eq.~(\ref{eqn:rhoxContS}).  The next two conditions
represent the continuity of the phase, Eq.~(\ref{eqn:phiCont}),
and the derivative of the phase, Eq.~(\ref{eqn:alphaCont}).
Equation~(\ref{eqn:muCont}) demands that the eigenvalue be the
same on either side of the boundary.  Note that in
Eq.~(\ref{eqn:mu}) the eigenvalue on the right hand side is offset
by $V_0$ as compared to the left hand side.  Since $n$ only enters
into the phase and does not effect such quantities as the
eigenvalue and density, only the $n=0$ state is considered and,
consequently, the phase, $\phi$, is continuous across the
boundary.  However, it is important to note that all solutions are
of denumerably infinite degeneracy, according to
Eq.~(\ref{eqn:phiCont}).  Since $\phi$ is given by \be
\phi(x)=\int_0^x\frac{\alpha}{\rho(x)}\dd x+\mathrm{const.}\, ,
\ee
 continuity in the phase is easily achieved by setting the constant phase shift equal on either side of the boundary and, therefore, Eq.~(\ref{eqn:phiCont}) is satisfied.

In the following derivation, it is assumed that the wavefunction
parameters on the left side of the step are known completely.
Therefore the density prefactor, $A_L$, the vertical density
offset, $B_L$, the translational scaling, $b_L$, and the
horizontal offset, $x_{0L}$ are all known, where the $L$ subscribt
refers to variables on the left side; an $R$ subscript will refer
to variables on the right side.  In addition, the experimental
parameters of the interaction strength, $g$, and the potential
step height, $V_0$, are both known.  From the variables on the
left and Eq.~(\ref{eqn:sn}), the density at the boundary, $\rho_L
\equiv \rho(0^-)$, and its derivative at the boundary, $\partial_x
\rho_L \equiv \partial_x \rho(0^-)$, can be determined.  The
eigenvalue, $\mu$, and the phase constant, $\alpha$, can be
determined from Eqs.~(\ref{eqn:mu}) and~(\ref{eqn:alpha}).

From Eqs.~(\ref{eqn:sn}) and~(\ref{eqn:rhoCont}) the square of the
Jacobian elliptic $sn$ function can be solved for, \be \sn^2(x_{0
R},\sqrt{\frac{A_R g}{b_R^2}})=\frac{\rho_L-B_R}{A_R}\, ,
\label{eqn:snr} \ee and, from Eqs.~(\ref{eqn:mu})
and~(\ref{eqn:muCont}), the horizontal scaling, $b_R$, \be
b_R^2=2(\mu-V_0)-(A_R+3 B_R)g\, . \label{eqn:br} \ee These
variables are substituted into Eqs.~(\ref{eqn:alphaCont})
and~(\ref{eqn:rhoxContS}), using Eq.~(\ref{eqn:alpha}), to give
\bea
\alpha_L^2&=&B_R(A_R+B_R)\nonumber\\&&\times[2(\mu-V_0)-(A_R+2B_R)g]\,
, \label{eqn:alphaLS}
\\
(\partial_x
\rho_L)^2&=&-4(B_R-\rho_L)(A_R+B_R-\rho_L)\nonumber\\&&\times[2(\mu-V_0)-(A_R+2B_R+\rho_L)g]\,
. \label{eqn:rhoLS} \eea Equations~(\ref{eqn:alphaLS})
and~(\ref{eqn:rhoLS}) are quadratic in $A_R$ and cubic in $B_R$
and can be solved analytically for $A_R$ and $B_R$ to give six
solutions.  The remaining variables on the right side can then be
found by substituting the values of $A_R$ and $B_R$ into
Eq.~(\ref{eqn:br}), to find $b_R$, and by taking the inverse
Jacobi $sn$ function of Eq.~(\ref{eqn:snr}) to give \be
x_{0R}=\sn^{-1}(\frac{\rho_L-B_R}{A_R},\sqrt{\frac{A_R
g}{b_R^2}})\, . \label{eqn:x0r} \ee The full solution is then
completely known.  It is therefore possible to completely describe
the system analytically knowing only the parameters on one side of
the step.  This not only introduces computational ease in
evaluating solutions, it also provides all possible solutions,
most of which cannot be determined using purely numerical methods.
In the following section, specific examples of a BEC in the
presence of a step potential will be examined.


\subsection{Particular Examples}
\label{subsec:StepExamples}

The solutions to the potential step problem can be divided into
two categories.  The eigenvalue, $\mu$, can be large enough that
particles are free to move across the boundary and the nonlinear
analog to a transmitted wave for the linear Schr\"odinger equation
becomes possible.  However, if the  eigenvalue is too small, then
the wavefunction must decay under the step.

\begin{figure}[tb]
\begin{center}
\epsfxsize=7.8cm \leavevmode \epsfbox{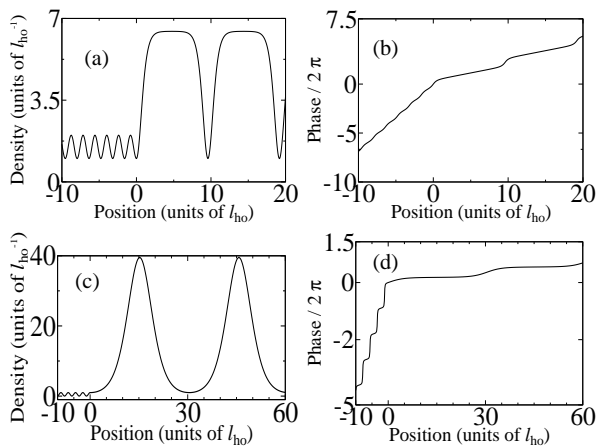}
\caption{Stationary solutions to the NLS  with a potential step of
form $V(x)=V_0\theta(x)$.  These solutions, which are the
nonlinear analogs of the continuum of linear stationary scattering
states, exhibit a large deviation from the traditional linear
solutions.   Shown are particular examples of (a) the density and
(b) the phase for a repulsive interaction strength and (c) the
density and (d) the phase for attractive interaction strength.
\label{fig:StepTrans}}
\end{center}
\end{figure}

 When $\mu$ is larger than the effective potential,
\be V_{eff}(x)=V_0 + g \rho(x)\, , \ee the wave can be transmitted
across the boundary.  Note, in the case of an attractive
interaction, $g < 0$, the eigenvalue can be less than the step
height, $V_0$, and for a repulsive interaction, $g > 0$, the
eigenvalue must be strictly greater than the step height.  In
Fig.~\ref{fig:StepTrans}(a) the density of a nonlinear state with
a repulsive interaction strength is shown.  A step of height
$V_0=1$, positioned at $x=0$, and a condensate with an interaction
strength of $g=0.2$ and eigenvalue of $\mu=2.404$ were used.
Notice that the increased interaction strength and nonlinearity
has caused the peaks of the wavefunction to become much broader
than in the linear case.  The phase that corresponds to this
density is shown in Fig.~\ref{fig:StepTrans}(b).
Figure~\ref{fig:StepTrans}(c) shows a similar solution but with an
attractive interaction strength.  This potential is again given by
a step with height of $V_0=1$, positioned at $x=0$.  An
interactionstrength of $g=-0.2$ and eigenvalue of $\mu=0.98$ was
used.  In this case, the peaks have instead narrowed due to the
attractive interaction.  The phase that corresponds to this
density is shown in Fig.~\ref{fig:StepTrans}(d).

%

\begin{figure}[tb]
\begin{center}
\epsfxsize=7.8cm \leavevmode \epsfbox{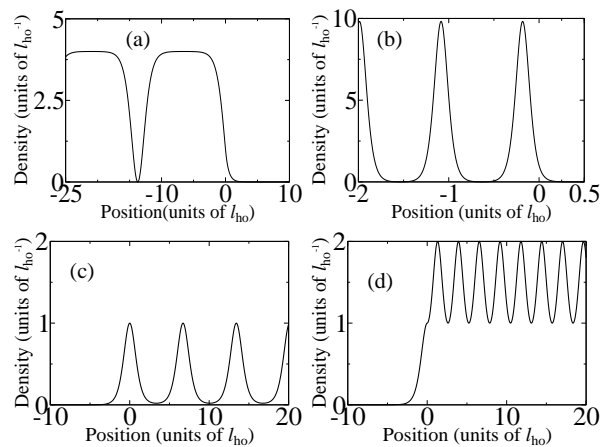}
\caption{Stationary solutions to the NLS with a potential step for
which the wavefunctions decay, which are the nonlinear analogs to
evanescent waves.  Shown are particular examples of the densities
of nonlinear waves with (a) repulsive interaction strength, (b)
attractive interaction strength, and (c) and (d) with an
attractive interaction strength that decay on the \emph{lower}
side of the step. \label{fig:StepDecay}}
\end{center}
\end{figure}

When the eigenvalue is less than the effective potential, $\mu <
V_{eff}(x)$, the wavefunction must decay  under the step.  In
Fig.~\ref{fig:StepDecay}(a) the density of a nonlinear solution
with a repulsive interaction strength that decays as it crosses
the boundary of the step is shown.   A step with height of
$V_0=1$, positioned at $x=0$, and a condensate with an interaction
strength of $g=0.12$ and eigenvalue of $\mu=0.5$ were used.
Figure~\ref{fig:StepDecay}(b) shows a similar solution but with an
attractive interaction strength.  This potential is again given by
a step with height of $V_0=1$, positioned at $x=0$.  An
interaction strength of $g=-10$ and eigenvalue of $\mu=-49$ were
used.  For both wave functions, the phase is necessarily trivial,
since all wavefunctions that approach zero at infinity must have
$A=-B$, and hence from Eqs.~(\ref{eqn:phix})
and~(\ref{eqn:alpha}), the phase is constant.  In order to stress
the importance of the effective potential, and not just the step
potential, Figs.~\ref{fig:StepDecay}(c) and (d) show the density
of two nonlinear wavefunctions with attractive interactions,
$g=-1$, that decay on the \emph{lower} side of the potential. Both
wavefunctions have an eigenvalue of $\mu=-0.5$ and a potential of
height of $V_0=0.01$ and $V_0=1$ were used for
Fig.~(\ref{fig:StepDecay}(c) and Fig.~\ref{fig:StepDecay}(d),
respectively.

%

Thus the NLS with a potential step has solutions to the step
potential that provide a connection between the linear solutions
and a wide variety of exotic nonlinear wavefunctions, as shall be
discussed in Sec.~\ref{sec:linear}.


\section{Point-like Impurity}
\label{sec:Delta}

In this section, the case of a potential of form \be
V(x)=V_0\,\delta(x)\, \ee is considered.  Such a potential models
an impurity which deforms the constant background potential on a
length scale much less than that of the healing length.  Given the
state of the system on the negative $x$, or left side of the
impurity, the variables on the right side of the impurity are
determined.  A negative value of $V_0$ represents an attractive
impurity, such as due to defects in hydrogen-bonded
chains~\cite{Kivshar1991,Kivshar1991_2}, and a positive value of
$V_0$ represents a repulsive impurity, such as with helium atoms
in a BEC~\cite{Capuzzi2000,Harms1998}.


\subsection{General Solution}
\label{subsec:DeltaGeneral}

The boundary conditions for an impurity are similar to those for
the potential step, except that the derivative of the wavefunction
experiences a discontinuity at the boundary.  Therefore, it is
necessary that
Eqs.~(\ref{eqn:rhoCont}),~(\ref{eqn:muCont}),~(\ref{eqn:alphaCont})
and~(\ref{eqn:phiCont}) must still be satisfied, as well as \be
\partial_x \rho(0^+)-\partial_x \rho(0^-)=-4 \rho(0) V_0\, .
\ee

It is again assumed that all variables on the left side of the
impurity are known as well as the experimental parameters of
interaction strength, $g$, and impurity strength, $V_0$.  Using a
treatment that is exactly analogous to that for the step function,
all of the parameters on the right side, given those on the left,
are analytically determined.  The only difference is that in
Eqs.~(\ref{eqn:br}),~(\ref{eqn:alphaLS}),~(\ref{eqn:rhoLS}),
and~(\ref{eqn:x0r}), the quantity $\partial_x \rho_L$ must be
replaced with $(\partial_x \rho_L-4V_0\rho_L)$ and $(\mu-V_0)$
must be replaced with $\mu$.  It is therefore possible to
completely describe the system analytically knowing only the
parameters on one side of the impurity.  In the following section,
examples of a wavefunction subject to an impurity are examined.


\subsection{Particular Examples}
\label{subsec:DeltaExamples}

For the delta function potential, both symmetric and nonsymmetric
wavefunctions are possible.  Of particular interest are the
symmetric wavefunctions in the $k=1$ limit of the Jacobian
elliptic functions.  In this case all solutions become hyperbolic
trigonometric functions with a localized change in the density
around the impurity and no oscillations at $\pm \infty$. Solutions
of this type we term {\it localized}~\cite{Carr2001_2}. Due to the
form of the solutions, there are four different solution types.
The possible wave functions are then \bea
\rho&=&\alpha^2+(1-\alpha^2)\tanh^2(\sqrt{1-\alpha^2}|x|+x_0)\, ,
\label{eqn:tanh}
\\
\rho&=&\alpha^2+(1-\alpha^2)\coth^2(\sqrt{1-\alpha^2}|x|+x_0)\, ,
\label{eqn:coth}
\\
\rho&=&b^2 \sech^2(b |x|+x_0)\, , \label{eqn:sech}
\\
\rho&=&b^2 \csch^2(b |x|+x_0)\, , \label{eqn:csch} \eea where the
translational offset, $x_0$, is determined by the impurity
strength, $V_0$, and the density has been normalized according to
Eq.~(\ref{eqn:norm}).  Equations~(\ref{eqn:tanh})
and~(\ref{eqn:coth}) are valid for repulsive interactions, while
Eqs.~(\ref{eqn:sech}) and~(\ref{eqn:csch}) are valid for
attractive interactions.

\begin{figure}[tb]
\begin{center}
\epsfxsize=7.8cm \leavevmode \epsfbox{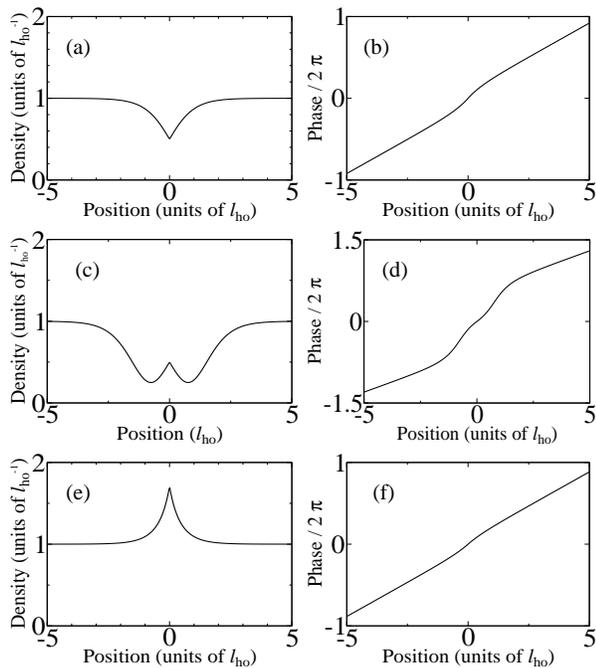}
\caption{Localized, symmetric solutions to the NLS with repulsive
interaction strength in the presence of an impurity, $V(x)=V_0
\delta (x)$.  Shown are particular examples of (a) the density and
(b) the phase of a dark soliton bound by a repulsive impurity, (c)
the density and (d) the phase of a pair of dark solitons bound by
an attractive impurity, and (e) the density and (f) the phase of a
supercurrent deformed by an attractive impurity.  Note that (a)
and (b) may also be interpreted as deformations of a supercurrent.
\label{fig:DelSymRep}}
\end{center}
\end{figure}

Figure~\ref{fig:DelSymRep} shows the possible wavefunctions for
repulsive interactions.  An example of the solution described by
Eq.~(\ref{eqn:tanh}) is plotted in Fig.~\ref{fig:DelSymRep}(a)
with a repulsive impurity of strength $V_0=0.5$.  This may be
interpreted as a single dark soliton bound by an impurity.  This
is similar to the solution found by Hakim~\cite{Hakim1997} for a
soliton that is moving with an impurity.   In addition, a bound
state of two dark solitons, Fig.~\ref{fig:DelSymRep}(c), can be
created when the strength of the impurity is attractive and
exactly balances the repulsion between the two dark solitons.
Figure~\ref{fig:DelSymRep}(e) shows the hyperbolic cotangent
function solution with an impurity strength of $V_0=-0.5$; this
may be interpreted as a deformation of the ground state constant
solution to the NLS with a constant potential.  In all plots an
interaction strength of $g=1$ and phase constant of $\alpha=0.5$
were used.

It should be noted that there exists a bound state of a
repulsive condensate with an attractive impurity.  This solution
is given by, \be \frac{b^2}{g} \mathrm{csch}^2(b x +
\mathrm{coth}^{-1}(\frac{-V_0}{b}))\, , \ee where the interaction
strength, $g$ has specifically been included and $b$ must be
determined such that the density is normalized to unity.  These
requirements place a limit on how repulsive the interaction may
become and is given by, \be g_{\mathrm{max}}=-4 V_0\, , \ee where
$g_{\mathrm{max}}$ is the most repulsive interaction the
condensate may have.  If the interaction is increased past this
point, the condensate will spill away from the impurity and will
no longer be bound.

\begin{figure}[tb]
\begin{center}
\epsfxsize=7.8cm \leavevmode \epsfbox{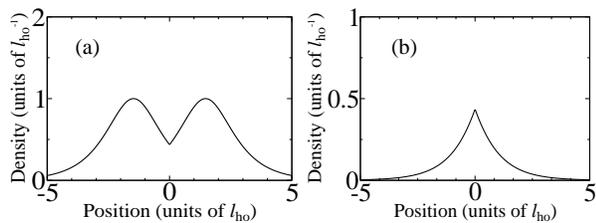}
\caption{Localized, symmetric solutions to the NLS with attractive
interaction strength in the presence of an impurity.  Shown are
particular examples of (a) the density of a bright soliton, which
is the ground state solution to the NLS, deformed by a repulsive
impurity and (b) the density of a bright soliton deformed by an
attractive impurity. \label{fig:DelSymAtt}}
\end{center}
\end{figure}

The set of symmetric localized solutions for the case of
attractive interactions do not allow for nontrivial phases, in
contrast to the case of repulsive interactions.  For attractive
interactions, $g<0$, the hyperbolic secant function solution,
Eq.~(\ref{eqn:sech}), is valid for both $V_0>0$,
Fig.~\ref{fig:DelSymAtt}(a), and $V_0<0$,
Fig.~\ref{fig:DelSymAtt}(b), where potential strengths of
$V_0=0.9$ and $V_0=-0.9$ were used, respectively.  These solutions
may be interpreted as a single bright soliton, which is the ground
state solution to the 1D-NLS, deformed by an impurity.  The
hyperbolic cosecant function solution, Eq.~(\ref{eqn:csch}), is
only valid if $V_0<0$ and is similar in form to the hyperbolic
secant solution of Fig.~\ref{fig:DelSymAtt}(b).  The two solutions
types are degenerate for $V_0<0$, with an eigenvalue of
$\mu=-b^2/2$.  In Figs.~\ref{fig:DelSymAtt}(a) and (b) an
interaction strength of $g=-1$ and translational scaling of $b=1$
were used.

%

%

\begin{figure}[tb]
\begin{center}
\epsfxsize=7.8cm \leavevmode \epsfbox{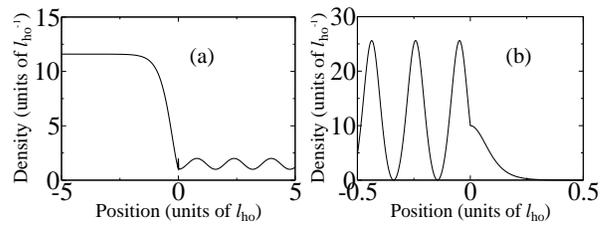}
\caption{Solutions to the NLS with an impurity that oscillate on
one side of the potential.  Shown are particular examples of (a)
the density of a nonlinear wave with repulsive interaction
strength and (b) the density of a nonlinear wave with attractive
interaction strength.  These waves have no analog with the
solutions of the linear Schr\"odinger equation.
\label{fig:DelHalfFigs}}
\end{center}
\end{figure}

Nonsymmetric wavefunctions which oscillate at infinity are also
possible and come in two forms: oscillations on one side of the
delta function and oscillations on both sides.
Figure~\ref{fig:DelHalfFigs} shows two possible nonsymmetric
wavefunctions subject to a delta function positioned at $x=0$ that
oscillate on one side of the potential.  The density of a
wavefunction with a repulsive interaction strength of $g=0.21$ and
eigenvalue of $\mu=2.4$, distorted by a delta function, $V_0=2$,
is shown in Fig.~\ref{fig:DelHalfFigs}(a), where the left side
reproduces the hyperbolic tangent function of
Eq.~(\ref{eqn:tanh}).  In Fig.~\ref{fig:DelHalfFigs}(b), an
attractively interacting, $g=-50$, wavefunction with an eigenvalue
of $\mu=-50$, distorted by a delta function, $V_0=10$, is shown
that appears similar to the evanescent wavefunctions of
Fig.~\ref{fig:StepDecay} that decay beneath a step.  Note that
Fig.~\ref{fig:DelHalfFigs}(a) has a nontrivial phase while the
phase of Fig.~\ref{fig:DelHalfFigs}(b) is trivial.

%

\begin{figure}[tb]
\begin{center}
\epsfxsize=7.8cm \leavevmode \epsfbox{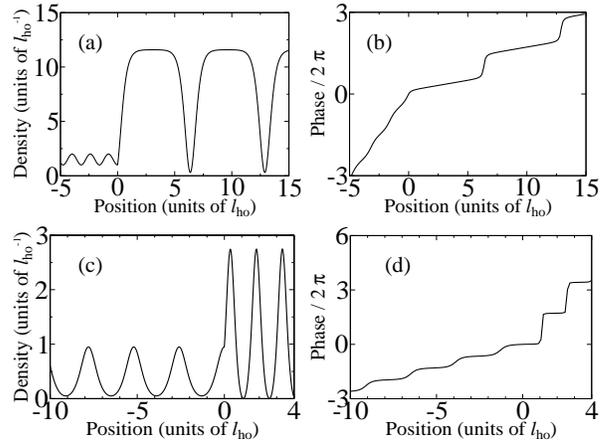}
\caption{Nonsymmetric solutions to the NLS with an impurity. Shown
are particular examples of (a) the density and (b) the phase of a
nonlinear wave with repulsive interaction strength and (c) the
density and (d) the phase of a nonlinear wave with attractive
interaction strength.  These solutions are the nonlinear analogs
to the continuum of linear stationary scattering states.
\label{fig:DelASymFigs}}
\end{center}
\end{figure}

Figure~\ref{fig:DelASymFigs} shows two possible nonsymmetric
wavefunctions subject to a delta function positioned at $x=0$ with
strength $V_0=2$ that oscillate on both sides of the potential.  A
repulsive interaction strength produces the characteristic
widening of the pulse peaks, Fig.~\ref{fig:DelASymFigs}(a).  The
corresponding phase is given in Fig.~\ref{fig:DelASymFigs}(b).  An
attractive interaction strength creates a narrowing of the pulse
peaks, Fig.~\ref{fig:DelASymFigs}(c).   The corresponding phase is
given in Fig.~\ref{fig:DelASymFigs}(d).  The condensates in
Figs.~\ref{fig:DelASymFigs}(a) and (c) are characterized by
eigenvalues of $\mu=2.4$ and $\mu=-1.3$, respectively.

%


\section{Linear Limits}
\label{sec:linear}

In this section it is shown how the solutions to the NLS connect
to the solutions of the linear Schr\"odinger equation.  There are
two distinct types of possible waves.  If the wave has enough
energy it is possible to make the wave propagate through space.
However, if it does not have enough energy, the wave can carry no
current and is refered to as an evanescent wave.  These waves must
decay.  In the next two sections, these two types of waves are
discussed concerning their role as the linear limit of the
nonlinear solutions.


\subsection{ Transmitted Waves}
\label{subsec:linearTrans}

In a linear system, if the energy of the system is greater than
the potential energy, then the wave can be transmitted through
space as sine waves.  The nonlinear Jacobian elliptic {\it sn}
function waves are the nonlinear analogue of the sine waves of a
linear system.  In this section, it is shown how the nonlinear
solutions, in the linear limit, recreate the linear solutions of
propagating waves.

For the linear case, the usual representation of the system by
incident, transmitted, and reflected waves is given by, \bea
\Psi_L(x,t)&=&\left(e^{i k_L x}+R e^{-i k_L x}\right)\, e^{-i\mu
t}\, ,
\\
\Psi_R(x,t)&=&T e^{i k_R x}\,e^{-i\mu t}\, , \eea where
$k_L=\sqrt{2 (\mu-V_L)}$, $k_R=\sqrt{2 (\mu-V_R)}$, $\mu$ is the
eigenvalue of the Schr\"odinger equation, $R$ is the reflection
coefficient, $T$ is the transmission coefficient, and the incident
wave is assumed to be coming in from the left.   The potentials,
$V_L$ and $V_R$, and the scattering coefficients, $R$ and $T$, are
determined by the type of boundary.  For the case of a potential
step, the potentials are given by $V_L=0$ and $V_R=V_0$.  For a
delta function potential, the potentials are given by $V_L=V_R=0$.
The wavefunction can be alternately described by an amplitude and
phase as follows: \bea
\rho_L&=&(1+r)^2-4\,r\,\sin^2(\sqrt{2\,\mu}\,x-s/2)\, ,
\\
\partial_x \phi_L&=&\frac{(r-1)(r+1)\sqrt{2\,\mu}}{(1+r)^2-4\,r\,\sin^2(\sqrt{2\,\mu}\,x-s/2)}\, ,
\eea where \bea
\Psi_L(x,t)&=&\sqrt{\rho_L(x)}\,\exp[i\phi_L(x)]\,e^{-i\mu t}\, ,\\
R&=&r e^{i s}\, , \eea with $r$ and $s$ real.  The nonlinear
solution is then connected to the linear solution by \bea
x_0&=&-s/2\, , \label{eqn:Stepx0}
\\
b^2&=&2\mu\, , \label{eqn:Stepmu}
\\
A&=&-4r\, , \label{eqn:StepA}
\\
B&=&(r+1)^2\, . \label{eqn:StepB} \eea
 The transmitted wave can easily be given by
\bea \rho_R&=&t^2\, ,
\\
\partial_x \phi_R&=&\frac{\sqrt{2 \mu}}{t^2}\, ,
\eea where \bea
\Psi_R(x,t)&=&\sqrt{\rho_R(x)}\,\exp[i\phi_R(x)]\,e^{-i\mu t}\, ,\\
T&=&t e^{i w}\, , \eea with $t$ and $w$ real.  The nonlinear
solution is then connected to the linear solution by \bea
x_0&=&0\, ,
\\
b^2&=&2\mu\, ,
\\
A&=&0\, ,
\\
B&=&t^2\, . \eea The nonlinear solutions when the eigenvalue is
greater than the effective potential are therefore adiabatically
connected to the linear transmitted wave solutions.  The next
section connects the decaying evanescent waves with nonlinear
solutions.


\subsection{ Evanescent Waves}
\label{subsec:linearDecay}

If the eigenvalue, $\mu$, is less than the effective potential
then the wave function must decay.  In the linear case, the decay
is precisely exponential.  The Jacobian elliptic $sn$ function
solution, Eq.~(\ref{eqn:sn}), of the NLS can provide the
appropriate exponential decay.  In the limits $g\rightarrow 0$ and
$B,\,x_0 \rightarrow +\infty$ under the constraints $A$=$-B$ and
$A\,g/b^2=1$, Eq.~(\ref{eqn:sn}) gives \be
\lim_{\substack{B,x_0\to+\infty}}B\,\sech^2(b x+x_0)= \rho_0
\,e^{-b x}\, , \ee
 where $\rho_0$ is the density at the boundary and is given by,
\be \rho_0=\frac{4 B}{e^{2 x_0}}\, . \ee When the limit that $g$
approach zero is not enforced, \be \rho(x)\propto
\frac{1}{(N_+\exp^{+b x}+N_-\exp^{-b x})^2}\, , \ee where $N_+$
and $N_-$ are constants related to the magnitude and sign of the
interaction strength.  The decay of the density is, therefore, not
strictly an exponential decay.  It is interesting to note that all
decaying solutions, whether linear or nonlinear, must possess a
trivial phase.  This is due to the restriction that $A$=$-B$, and,
hence from Eqs.~(\ref{eqn:alpha}) and~(\ref{eqn:phix}), the phase
must vanish.  This is also consistent with the physical
interpretation of the phase since a nontrivial phase corresponds
to a superfluid velocity and the velocity must vanish if the wave
cannot be transmitted.


\section{Discussion and Conclusions}
\label{sec:conclusion}

The full set of stationary states of the mean field of a
Bose-Einstein condensate, modeled by the nonlinear Schr\"odinger
equation in one dimension, in the presence of a potential step or
pointlike impurity were presented in closed analytic form.
Non-decaying solutions were divided into two categories: localized
soliton-like solutions, and solutions that oscillate out to
infinity.  The localized solutions are of a purely nonlinear
character, as they have no linear analog.  The oscillating
solutions, on the other hand, were shown to be adiabatically
connected to the solutions to the linear Schr\"odinger equation.

The localized solutions present novel wavefunctions.  With a
delta function potential, the localized solution can be
interpreted as a single bright or dark soliton trapped by the
impurity.  In addition, it was shown that an impurity can also
bind a soliton pair.  If the impurity is attractive, the natural
repulsion between two dark solitons can be exactly canceled by the
attraction of the impurity, while if the impurity is repulsive, it
can balance the natural attraction of in-phase bright solitons.
Since these solutions are conjected to be stable (see below), they
are excellent candidates for the experimental realization of
stationary excited states of a Bose-Einstein condensate.  In
addition, the maximum repulsive interaction strength of the
condensate with an attractive impurity that allows for a bound
state has been determined.

The oscillating solutions to the NLS, despite being adiabatically
connected to oscillating solutions to the Schr\"odinger equation,
have very different properties due to the concept of an effective
potential.  For the attractive interaction solution for an
evanescent wave decaying under a step, as illustrated in
Fig.~\ref{fig:StepDecay}(b), the eigenvalue is larger than the
effective potential in the regions of high density and is less
than the effective potential in regions of low density.
Figures~\ref{fig:StepDecay}(c) and (d) show more radical
deviations from the linear solutions since the wavefunctions decay
on the lower side of the step.

It is possible to characterize the general solution to the NLS,
Eq.~(\ref{eqn:sn}), in terms of physical quantities such as the
mean linear number density, $\overline{n}$ , mean energy density,
$\mathcal{\overline{E}}$, and mean momentum density,
$\mathcal{\overline{P}}$.  The densities are given by \bea
\overline{n}&=&B + A \Big(\frac{1}{k^2}-\frac{E(k)}{k^2
K(k)}\Big)\, , \label{eqn:number}
\\
\mathcal{\overline{E}}&=&\overline{n} \mu + \frac{3 B^2 g - b^2
A}{6}+\frac{2}{3}(\mu-V)(\overline{n}-B)\, , \label{eqn:energy}
\\
\mathcal{\overline{P}}&=&\alpha\, , \label{eqn:momentum} \eea
where $\mu$, $\alpha$, and $k$ are given by
Eqs.~(\ref{eqn:mu}),~(\ref{eqn:alpha}), and~(\ref{eqn:m}),
respectively.   In general, the number, energy, and momentum
densities must be calculated separately for the left and right
sides of a boundary.  These densities can be used to determine the
variables $A$, $B$, and $b$ of Eq.~(\ref{eqn:sn}), leaving only
the translational offset, $x_0$, as a free variable determined by
the boundary conditions.  It should be noted that the factor
multiplying $A$ in the mean number density,
Eq.~(\ref{eqn:number}), approaches one half when $k$ approaches
zero and approaches one when $k$ approaches unity.  This is to be
expected since the mean number density of a linear wave is given
by $B+A/2$ and the mean number density of an extremely nonlinear
wave is given by $B+A$.  The mean energy density can easily be
calculated from \be \mathcal{\overline{E}} = \overline{n} \mu -
\frac{g}{2}\overline{\rho^2}\, , \ee and so the second and third
terms on the right side of Eq.~(\ref{eqn:energy}) are due to this
nonlinear correction.  Since the mean momentum density,
Eq.~(\ref{eqn:momentum}), is equal to $\alpha$, the mean momentum
density must be equal on both sides of the boundary due to the
boundary condition on $\alpha$, Eq.~(\ref{eqn:alphaCont}).  In
addition, the momentum density as a function of position is also
given by $\alpha$ and so the momentum density is equal everywhere.

The healing length, $\xi$, of the NLS in the quasi-one-dimensional
regime, where the transverse dimensions are trapped by a harmonic
potential of frequency $\omega$, is given by \be
\xi^2=\frac{l_{ho}^2}{8 \pi a_s \overline{n}}, \ee where
$\overline{n}$ is given by Eq.~(\ref{eqn:number}).  Since the mean
number density can vary across the boundary, it is possible for
the condensate to have a different healing length on either side
of a boundary.  Since the speed of sound in the condensate is
inversely proportional to the healing length, the speed with which
phonon-like excitations can travel vary as they cross the
boundary.

While finding the complete set of solutions to
Eq.~(\ref{eqn:NLSE}) with an impurity or step potential provides
much information about the system, only stable solutions are
experimentally observable. Previous works have examined the
stability of stationary states for a constant external potential
(see, for
example,~\cite{Infeld2000,Infeld1979,Gordon1983,Gordon1986,Haus1996,Kivshar1998,Sulem1999,Yuen1975}),
as well as for periodic and harmonic potentials (see, for
example,~\cite{Bronski2001_2,Carr2001_3}). Most studies are
ultimately numerical: linear stability can be solved in a few
special cases, while nonlinear stability is analytically
intractable. For a constant external potential, single bright and
dark solitons and dark soliton trains are stable. A finite number
of bright solitons may form bound states, as for example order $n$
solitons ($n>1$). Bright soliton trains are always unstable, but
may be experimentally stable, in that their lifetime is much
longer than experiments, which typically require stability
timescales of from milliseconds to seconds. Bright soliton trains
which have a phase difference $\Delta\phi$ between adjacent peaks
such that $-\pi/2 < \Delta\phi < 3\pi/2$ exhibit this experimental
stability, with the lifetime being longer the closer $\Delta\phi$
is to $\pi$. Bright soliton trains with $-\pi/2 < \Delta\phi <
\pi/2$ are unstable but become quasi-periodic in time in a finite
system.

Based on these known results from the case of a constant
potential, as well as the from the stability analysis with an impurity performed by Bogdan {\it et al.}~\cite{Bogdan1997}, the stability of an attractive condensate with an impurity is as follows.  According to Bogdan, the bound state of two bright solitons, as in
Fig.~\ref{fig:DelSymAtt}(b), are stable, since, so long as
they are strongly overlapping, they will be in phase
($\Delta\phi=0$) and remain bound to the impurity. This is also the ground state of the system.  However, the kind of solution shown in Fig.~\ref{fig:DelSymAtt}(a) is unstable~\cite{Bogdan1997}.  We then conject on the stability of a repulsive condensate that is not bound and whose density approaches a nonzero constant at infinity.  Localized solutions
in the case of repulsive nonlinearity are obviously stable in the
cases of Fig.~\ref{fig:DelSymRep}(a)-(b)
and~\ref{fig:DelSymRep}(e)-(f), since they are the ground state.
The bound pair of dark solitons, illustrated in
Fig.~\ref{fig:DelSymRep}(c)-(d), should be likewise stable, so
long as the impurity is sufficiently strong.  All of these
solution types, except for attractive solitons bound by a repulsive imputiry, are expected to be experimentally observable in
finite systems, such as an elongated harmonic trap.  An excellent analysis of the stability of solitons pinned with impurities is given by Bogdan, {\it et. al.}~\cite{Bogdan1997}.

The stability of soliton trains is a less certain issue.  In the
repulsive case, the central question is whether or not the phase
locking of the individual solitons in the train is destroyed by
the impurity or potential step; if it is, they may become unstable
in the region of the discontinuity in the potential.  In the case
of bright soliton trains, unless there is a strong phase
difference between the peaks, they will attract and become
unstable; otherwise, the discontinuity should not present a source
of instability, since bright solitons adjust themselves to
perturbation by emission of a small fraction of the total
wavefunction~\cite{Haus1996}. In order to perform numerical
studies, the solutions would have to be quantized on a ring, in
order to provide a finite domain for simulation.  Such a stability
study presents a subject for future research.

We emphasize that neither the idea of left and right traveling
waves nor that of reflected {\it plus} incident waves apply to
nonlinear wave equations.  This is important since one cannot
create wavepackets from linear combinations of these solutions.
Instead, these solutions already contain the wavepacket-like
solutions, or solitons, that are necessary to describe the system;
moreover, solitons, unlike wavepackets, are nondispersive.  Time
dependent nonlinear scattering remains an open question that can
certainly be addressed via numerical studies. In general, the
stationary solutions to the NLS give physical insight into its
dynamics, without which numerical solutions may be difficult to
interpret.  Perhaps more importantly, using the general nature of
the solutions to the cases of a step function and an impurity, it
is possible to describe all stationary states to piecewise
constant potentials.

In conclusion, we have analytically solved for all stationary
solutions to the nonlinear Schr\"odinger equation with a delta
function or a step function potential.  This models the steady
state behavior of the mean field of a Bose-Einstein condensate in
the presence of an impurity, or of a potential step created by,
for instance, a laser passing over the edge of a razor blade.
Novel wavefunctions were found, including solitons trapped by the
impurity and the nonlinear analog of transmitted and evanescent
waves.


\begin{center}
{\bf Acknowledgments}
\end{center}

We acknowledge helpful discussions with J. Cooper. Support is
acknowledged for B.T.S. from the National Science Foundation and
for L.D.C. from the U.S. Department of Energy, Office of Basic
Energy Sciences via the Chemical Sciences, Geosciences and
Biosciences Division, as well as the National Science Foundation
via grant no.~MPS-DRF 0104447.

\appendix


\section{Jacobian Elliptic Functions}
\label{app:jacobi}

A brief review of the Jacobian elliptic
functions~\cite{Abramowitz1964,Milne1950} is given.  Of the 12
elliptic functions, there are only six that are normalizable.  Of
these six, only three represent a different physical form, $sn$,
$cn$ and $dn$.  However, they are still related by, \bea
\cn^2&=&1-\sn^2\, ,
\\
\dn^2&=&1-k^2\,\sn^2\, . \eea The six non-normalizable elliptic
functions can also be reduced through a phase shift to three with
different forms, $ns$, $ds$, and $cs$, which are also related by,
\bea \cs^2&=&\ns^2-1\, ,
\\
\ds^2&=&\ns^2-k^2\, . \eea The normalizable and nonnormalizable
functions can be related through \be \sn^2(i K(1-k^2)+z,k)=k^2\
\ns^2(z,k)\, , \ee where $K(x)$ is the complete elliptic integral
of the first kind.  Therefore the square of any elliptic function
can be related linearly to $sn^2$.

The limits of the $sn$, $cn$ and $dn$ functions, along with the
complete elliptic integrals $K(k)$ and $E(k)$ are presented in
Table~\ref{table:jacobi}.  The period of the $sn^2$, $cn^2$ and
$dn^2$ functions is $2 K(k)$.
\begin{table}[!t]
\caption{Limits of the Jacobian elliptical functions and
integrals~\cite{Abramowitz1964}.} \label{table:jacobi}
\begin{tabular}{c c c}
\hline \hline
                &  $k=0$    & $k=1$       \\
\hline
$\sn(u,k)$      & $\sin(u)$ & $\tanh(u)$  \\
$\cn(u,k)$      & $\cos(u)$ & $\sech(u)$  \\
$\dn(u,k)$      & $1$       & $\sech(u)$  \\
$\ns(u,k)$      & $\csc(u)$ & $\coth(u)$  \\
$\ds(u,k)$      & $\csc(u)$ & $\csch(u)$  \\
$\cs(u,k)$      & $\cot(u)$ & $\csch(u)$  \\
$\mrm{K}(k)$    & $\pi/2$   & $\infty$    \\
$\mrm{E}(k)$    & $\pi/2$   & $1$         \\
\hline \hline
\end{tabular}
\end{table}


\section{Complete Solution Set}
\label{app:solution}

It is possible to prove that Eq.~(\ref{eqn:rhoInt}) is a Jacobian
elliptic integral of the first kind and can therefore be inverted
to produce the Jacobian elliptic functions.  If arbitrary
parameters are used, Eq.~(\ref{eqn:rhoInt}) becomes,
\be
\int \frac{1}{\sqrt{A_3 \rho^3+A_2 \rho^2+A_1 \rho+A_0}}\, \dd \rho =
x+x_0\, ,
\ee
where the $A_i$'s are real constants.  The cubic
polynomial can be factored to give,
\be
\int \frac{1}{\sqrt{(\rho+B_1)(\rho+B_2)(\rho+B_3)}}\, \dd \rho =
x+x_0\, ,
\label{eqn:rhoi}
\ee
where at least one of the constants $B_i$ must be real.  Without loss of generality we may take the
real constant as $B_1$.  The substitution $\rho=y^2-B_1$ is then
made in Eq.~(\ref{eqn:rhoi}) to yield \be \int
\frac{2}{\sqrt{(y^2+(B_2-B_1))(y^2+(B_3-B_1))}}\, \dd y = x+x_0\,
. \ee This is the general form of the elliptic integral of the
first kind~\cite{Abramowitz1964} and therefore gives \be C_1
\mrm{el}^{-1}(C_2 y,C_3)=x+x_0\, , \ee where the $C_i$'s are
constants and $el$ is one of the twelve elliptic functions.  This
can be inverted and $\rho$ replaced to produce
\be
\sqrt{\rho+B_1}=C_2^{-1} \mrm{el}(C_1^{-1} (x+x_0),C_3)\, ,
\ee
or
finally,
\be
\rho=C_2^{-2} \mrm{el}^2(C_1^{-1} (x+x_0),C_3)-B_1\, .
\ee
Since the square of any elliptic function can be related
linearly to the square of $sn$, only one independent solution of
the form \be \rho=A\, \sn^2(b x + x_0,k) + B\, , \ee need be
considered.

\bibliographystyle{prsty}

\end{document}